# Variations in Web of Science and Scopus Journal Coverage, Visibility and Prestige between 2001 and 2020

Toluwase Victor Asubiaro[2]

[2]African Research Visibility Initiative
Edmonton,
Canada
toluwase.asubiaro@africarvi.org, tasubiar@uwo.ca

**Abstract**:

*Purpose*: This study focuses on the changes in differences in the journal coverage, visibility and prestige of journals from top twenty countries in Web of Science and Scopus in the twenty-year timeframe-2001-2020.

*Methodology*: Using Web of Science and Scopus journal data from Journal Citation Reports and Scimago Journal Rank, respectively, top twenty countries by number of journals indexed in the two databases were identified. Analysis of the changes that occurred in the number of journals from the top twenty countries, the citations they received and their prestige were analyzed.

*Findings*: USA and UK continued their dominance of the journals indexed in Web of Science and Scopus, but their dominance waned gradually in the course of the twenty-year period. The rate of growth of journals indexed by the databases is steeper among the countries outside the top. In Web of Science, journals from the UK were the most prestigious until 2010 when China emerged as the most prestigious journals. USA continues to take the leading spot in terms of most prestigious journals in Scopus, followed by UK.

*Research Limitations*: This investigation relied on third-party datasets sourced from the Scimago Journal Rank repository for the compilation of the Scopus journal list.

*Practical implications*: This study suggests an inclination towards diversity by Web of Science and Scopus, though North America and Europe continue to dominate journal coverage. However, the gulf in the prestige and visibility of journals from North America, Europe and other parts of the world remains, suggesting the researchers from the *peripheral* may continue to gravitate towards the *core*.

*Originality/Value*: While studies have provided singular-year analyses of journal coverages of Web of Science and Scopus, this study provides an analysis of 20 year

1. Introduction

The influence of Web of Science and Scopus in the contemporary global research ecosystem is significant, serving as fundamental tools for research evaluation and ranking (Bohórquez, 2016; Pranckutė, 2021). These databases have the power to guide the trajectory of research, and journal being indexed in them is often synonymous with mainstream global science, while non-indexed journals can be relegated to the periphery (Beigel et al., 2023; Chavarro et al., 2017). Major global rankings and reports, including UNESCO's World Science Report, QS, and THE university rankings, and academic metrics like Scimago Journal Ranking and CWTS Leiden Ranking rely on Scopus and Web of Science data. These rankings significantly impact various facets of research evaluation, from university appointments and promotions to the allocation of research funding (Asubiaro & Onaolapo, 2023; Frenken et al., 2017; Haddawy et al., 2017). Despite alternative databases like Google Scholar, Microsoft Academic and Dimensions offering more comprehensive coverage, they have yet to match the influence wielded by Web of Science and Scopus in the policy and research landscape (Asubiaro et al., 2024). These databases continue to hold sway in research evaluation and policy decisions, influencing the global research landscape (Bohórquez, 2016).

In spite of the importance of Web of Science and Scopus, these databases are under scrutiny due to their overrepresentation of sources originating from Europe and North America (Asubiaro et al., 2024). The geographical locations of these databases in the global north, along with the significant influence of the "big five" academic publishing giants based in these regions, likely contribute to this bias. This geographical bias is a cause for concern as it may distort the representation of global academic knowledge, leaving research from other regions significantly underrepresented (Asubiaro et al., 2024; Larivière et al., 2015; Mongeon & Paul-Hus, 2016). These studies show the prominence of developed English-speaking countries and underscore the prominence of research from the United Kingdom and the United States (Mongeon & Paul-Hus, 2016; Singh et al., 2021). The historical dimension of the scrutiny is significant, with concerns about geographical and linguistic biases dating back to the 1970s when these issues were first raised regarding the Science Citation Index, now known as Web of Science. Despite early concerns, disputes arose, particularly from the database's creator, Garfield (Garfield, 1983, 1997; Goodwin & Garfield, 1980). Over the years, researchers revisited questions surrounding coverage and bias with critical scholars arguing that these databases, to some extent, obscure scientific contributions from so-called "third-world" regions and accentuate the influence of center-periphery dynamics in shaping coverage (Arunachalam & Garg, 1986; Aydinli & Mathews, 2000; Gibbs, 1995; Luukkonen et al., 1992; Van Leeuwen et al., 2001).

While studies have provided snapshot analysis of singular year highlights of journal coverages of Web of Science and Scopus, these studies do not provide years-long big picture of the changes in Web of Science and Scopus. A temporal analysis over a number of years is needed to understand the changes in the journal coverage in Web of Science and Scopus; specifically how the influence of the core countries has changed or the representation of sources from the periphery has alleviated. This study evaluates the changes in Web of Science and Scopus journal coverage of the top twenty countries within a twenty-year period (2001 to 2020).

## 2. Methodology

Web of Science journal lists for 2001 to 2020 years were downloaded from Web of Science Journal Citation Reports (JCR) portal. The source type was chosen as journals. Other source types like conference proceedings, book series and books were excluded. We excluded Crimea journal data because it was not recognized as a country or territory by the United Nations. Basically, sources like the Scimagor, ISSN portal, DOAJ, Worldcat and/or journal website were consulted to determine the journal country where journal country information was missing from data from Web of Science JCR portal. Journals where journal country information could not be decisively resolved were removed.

Scopus journal lists from 2001 to 2020 were downloaded from the Scimago Journal Ranking (SJR) Website[1]. The SJR metrics are based on Scopus data. Unlike Web of Science journal data, Scopus journal data from SJR was complete and only contained United Nations-recognized territories and countries. For consistency between data from Web of Science and Scopus, records of England, Scotland Wales and North Ireland were collated for the United Kingdom.

Web of Science and Scopus journal data for the twenty-year period included name of journal, country of publication, number of citations, and journal prestige (Impact Factor for Web of Science journals and SJR for Scopus journals) which were analyzed and presented in the results section. Journals' country of publication were ranked by the number of journals in Web of Science and Scopus. South Africa and Iran had the same number of journals in Web of Science in 2021. South Africa was chosen to occupy the twentieth and last spot because it had more journals than Iran in 2020. The analysis limited to the top twenty countries and an aggregated category labeled "OTHERS" represents countries outside the top twenty, not individually listed in the analysis.

Coverage: The number of journals indexed in a country per year represents the journal coverage of the country in Web of Science and Scopus.

Journal Share: The percentage share of each country's journal coverage in all the journals in Web of Science and Scopus was calculated to represent the journal share of a country.

The percentage increase in the number of journals year-over-year in a country was calculated to understand the how each country has increased/decreased its number of journals in the two databases.

Journal Visibility: The sum of the citations received by journals from a country was calculated to represent the visibility of the journals from the country.

Journal Prestige: The prestige of journals in Web of Science is the impact factor while Scimago Journal Rank represents journal prestige in Scopus. The average impact factor or Scimago Journal Rank of the journals in a country in a particular year represented the prestige of the journals in the country in the year. Scimago Journal Rank is a quantitative value (not a rank data type) that is calculated based on the prestige of the citing journals.

### 3. Results

The results of the twenty-year (2001-2020) analysis of journal coverage, visibility and prestige are presented under five headings. The first heading explores the number of journals indexed in Web of Science and Scopus from the top twenty countries. The second sub-section highlights the top twenty countries' share of the journals in Web of Science and Scopus. The third sub-section focused on the percentage changes in the top twenty countries journal coverage in Web of Science and Scopus in the twenty year period. The fourth sub-section harps on the top twenty countries' journals visibility while the last sub-section presents the result of the journal prestige.

---

[1] https://www.scimagojr.com/countryrank.php?order=itp&ord=desc&year=2021

*Top twenty Countries' Journal Coverage in Web of Science and Scopus from 2001 to 2020*

Figure 1 represents scientific journals published in various countries between 2001 and 2020. The top twenty countries in Web of Science in 2020 are USA (United States of America) , GBR (Great Britain), NLD (Netherlands), DEU (Germany), ESP (Spain), ITA (Italy), CHN (China), BRA (Brazil), RUS (Russia), FRA (France), POL (Poland), JPN (Japan), IND (India), CAN (Canada), KOR (South Korea), TUR (Turkey), AUS (Australia), CHE (Switzerland), COL (Colombia), ZAF (South Africa).

European countries dominated the top twenty positions in Web of Science journals, with nine countries—UK, Germany, the Netherlands, Spain, Switzerland, Italy, Russia, France, and Poland. Southeastern Asia followed with the second-highest number of countries, including Japan, China, and South Korea. North America (Canada and the United States) and Latin America (Colombia and Brazil) each had two countries in the top twenty, while Sub-Saharan Africa (South Africa), North Africa and West Asia (Turkey), Oceania (Australia), and Central and Southern Asia (India) each had one country represented.

Over the two-decade period, significant variations in growth rates in the number of journals indexed among countries by Web of Science were evident. The "OTHERS" category, encompassing countries outside the top twenty, exhibited the most substantial overall growth in the number of journals indexed in Web of Science, increasing from 514 in 2001 to 2796 in 2020—emerging only behind the USA and the UK. Spain notably ascended into the top five countries by 2005 from the bottom ten in 2001. The USA consistently maintained the highest number of journals indexed in Web of Science throughout the period, with moderate growth from 3,479 journals in 2001 to 5,943 journals in 2020. Great Britain demonstrated steady growth, positioning itself as a top publisher in Web of Science. Germany and the Netherlands consistently held their positions as top contributors.

In Europe, Spain and Switzerland witnessed the highest increase and rank change among the top twenty countries with journals indexed in Web of Science. In Latin America and the Caribbean, Colombia experienced a remarkable rise, increasing its journals from 1 to 72, while Brazil surged from 20 to 420 journals. Turkey and South Korea also exhibited drastic increases in their journal contributions.

Figure 1b illustrates the evolution of scientific journals published across various countries from 2001 to 2020. Several countries experienced significant differences in their growth during this period. The top twenty countries in Scopus in 2020 are USA (United States of America) , GBR (Great Britain), NLD (Netherlands), DEU (Germany), ESP (Spain), ITA (Italy), CHN (China), BRA (Brazil), RUS (Russia), FRA (France), POL (Poland), JPN (Japan), IND (India), CAN

(Canada), KOR (South Korea), TUR (Turkey), AUS (Australia), CHE (Switzerland), IRN (Iran) and CZE (Czechia Republic). Czechia Republic and Iran in Scopus replaced South Africa and Colombia in Web of Science in the top twenty list.

European countries continued their dominance in the Scopus top twenty positions for indexed journals, featuring ten nations—UK, Germany, the Netherlands, Spain, Switzerland, Italy, Russia, France, Czech Republic, and Poland. East and Southeastern Asia followed closely with three countries—Japan, China, and South Korea. North America (Canada and the United States) and Central and Southern Asia (India and Iran) each contributed two countries to the top twenty, while Latin America (Brazil), North Africa and West Asia (Turkey), and Oceania (Australia) each had one representative in the Scopus top twenty countries. Notably, no Sub-Saharan African country secured a position in the top twenty countries with journals indexed in Scopus. Japan was the sole nation among the top twenty that concluded the two-decade period with a reduction in the number of journals, decreasing from 490 in 2001 to 442 in 2020.

In contrast to Web of Science, the growth of the number of journals from both the top twenty countries and those outside the top twenty was generally sluggish in Scopus from 2001 to 2020. The USA consistently held the highest number of journals in Scopus during this period, with publications increasing from 5,062 papers in 2001 to 6,408 papers in 2020, indicating a steady but moderate ascent. Great Britain consistently expanded its presence and maintained its second-place position. The most notable growth, however, was observed in the number of journals indexed from countries outside the top twenty in Scopus. Iran's contribution saw a substantial increase, rising from only 7 journals in 2001 to 245 journals in 2020.

**Figure 1a: Top Twenty Countries with the highest Number of Journals in Web of Science between 2001 and 2020**

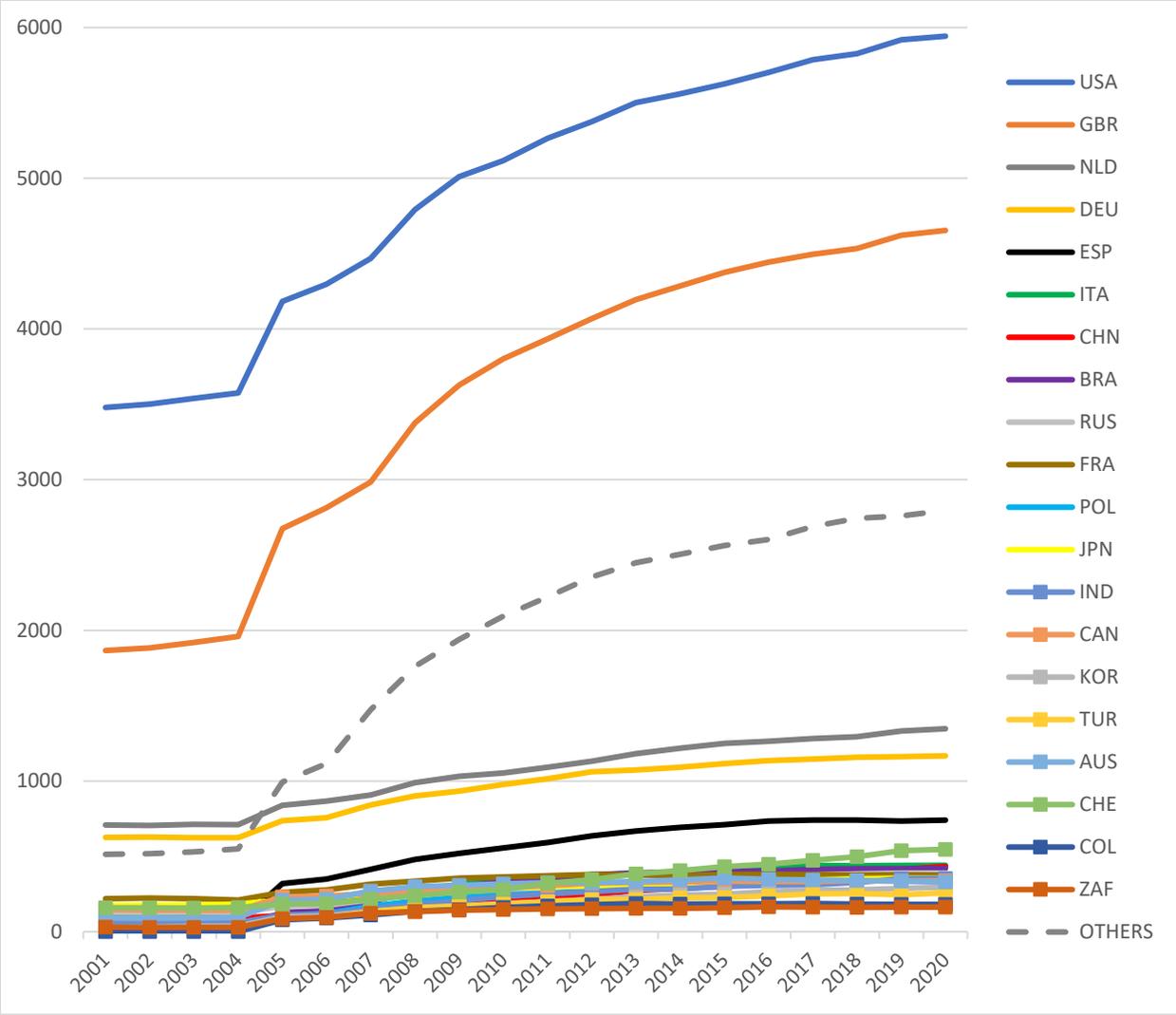

**Figure 1b: Top Twenty Countries with the highest Number of Journals in Scopus between 2001 and 2020**

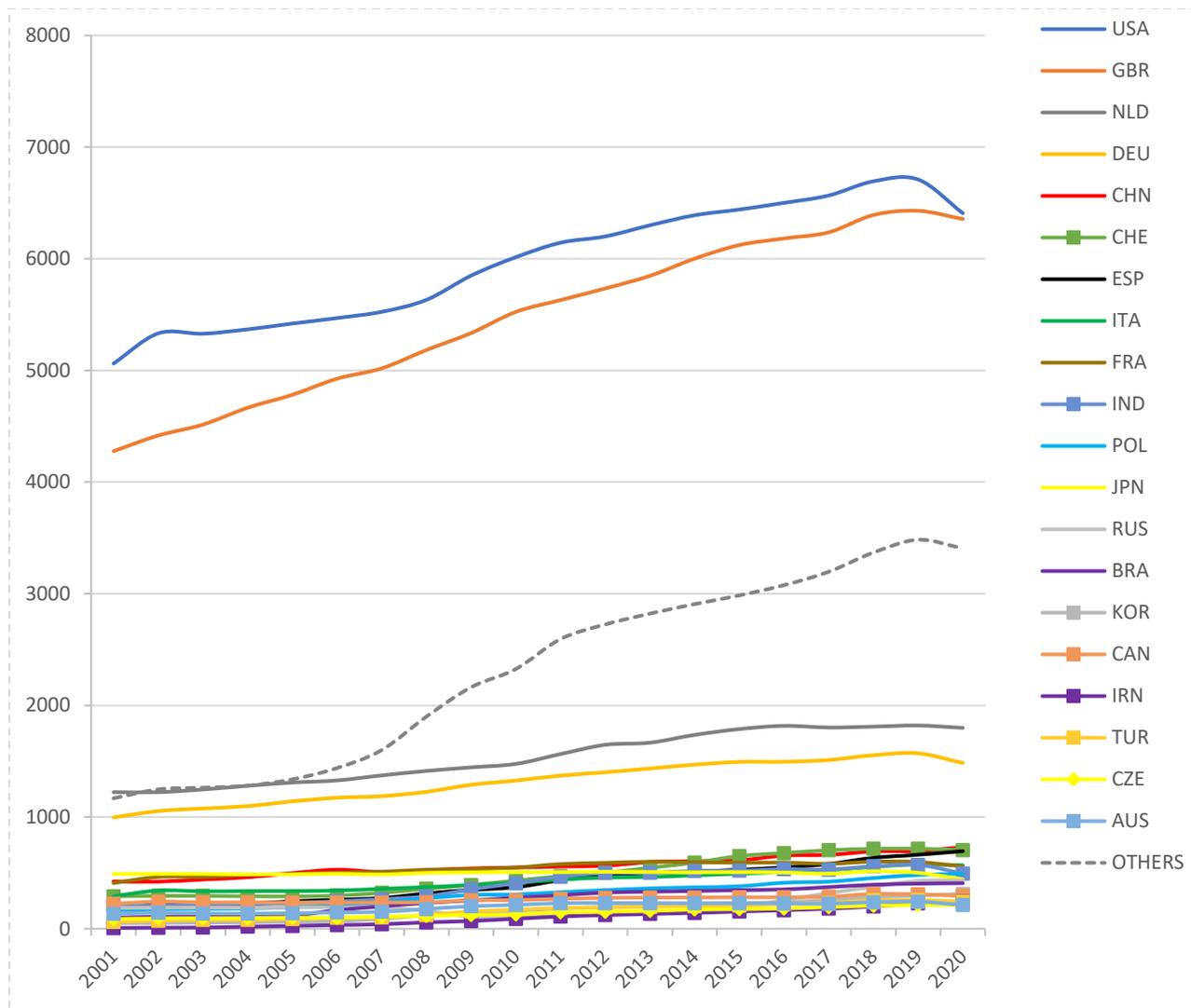

*Top twenty Countries' Journal Share in Web of Science and Scopus from 2001 to 2020*

Figure 2a illustrates the journal share of the top twenty countries in Web of Science journals globally from 2001 to 2020. The United States, the foremost contributor to Web of Science journals, witnessed a notable decline in its share, dropping from the initial 40.77% in 2001 to 27.16% in 2020. Similarly, Germany, France, Japan, and the Netherlands also experienced decreases in their shares of journals indexed in Web of Science. Conversely, the UK's share remained relatively stable at around 21% throughout the studied period. Notable increases in global journal shares were observed for Spain, Italy, Brazil, China, Poland, India, Korea, Turkey, South Africa, and Australia. Spain, for instance, saw a substantial rise from 0.56% in 2001 to 3.38% in 2020. Colombia, Turkey, Spain, and Korea exhibited the most significant increases in their shares. Despite a decline in most European countries, Spain stood out with an upward trajectory. In contrast, East and Southeastern Asian countries, specifically China and South Korea, increased their shares of journals in Web of Science over the two decades.

A discernible trend is the substantial increase in the share of Spanish-speaking countries, namely Spain and Colombia, in all journals indexed in Web of Science. Colombia, representing Latin America and the Caribbean, surged from a mere 0.01% in 2001 to 0.83% in 2020, while Brazil, also from the same region, increased its share from 0.23% to 1.92% within the same timeframe. Additionally, countries outside Europe and non-English-speaking regions, typically underrepresented in Web of Science, experienced an increase in their shares. Conversely, overrepresented countries, such as the UK, France, Germany, the Netherlands, and Japan, either reduced or remained largely unchanged in their global journal shares. The combined share of the USA and UK in Web of Science journals decreased from 62.62% in 2001 to 48.43% in 2020. The influence of the top four countries—USA, UK, Netherlands, and Germany—also diminished from 78.27% in 2001 to 59.92% in 2020, while the share of countries outside the top twenty increased from 6.02% to 12.78% in 2020.

Moving to Figure 2b, which presents data on the share of the top twenty countries in Scopus journals globally from 2001 to 2020, parallels with Web of Science are evident. The share of the United States in Scopus journals declined from 31.55% in 2001 to 24.18% in 2020, mirroring the trend observed in Web of Science. Likewise, top European countries, including the UK, Germany, France, Japan, and the Netherlands, experienced a drop in their shares. China and Australia's shares remained relatively constant over the two decades, while Spain, Italy, Brazil, Poland, India, Korea, Turkey, Iran, and Czechia Republic modestly increased their shares of global journals indexed in Scopus. The influence of the USA and UK in Scopus decreased from 81.39% in 2001 to 67.21% in 2020, and the influence of the top four countries declined from 72.05% in 2001 to 60.56% in 2020. Concurrently, the share of countries outside the top twenty increased from 7.28% in 2001 to 12.86% in 2020. Notably, one distinction between Web of Science and Scopus within the twenty-year period is the pace of change, with Web of Science experiencing a faster growth rate compared to the more gradual changes observed in Scopus.

**Figure 2a: Journal Share of the top Twenty Countries in Web of Science between 2001 and 2020**

**Figure 2b: Journal Share of the top Twenty Countries in Scopus between 2001 and 2020**

*Percentage changes in Country's Journal Coverage in Web of Science and Scopus between 2001 and 2020*

Table 1a, focuses on the academic journal publishing trends among the top twenty countries in Web of Science database between 2001 and 2020. The foremost observation that demands attention is the extraordinary growth in academic journal publications of some countries in the periphery such as Colombia and Turkey that increased their number of journals in Web of Science by 6360% and 18000% respectively between 2001 and 2020. Spain, and Latin American and the Caribbean countries -Brazil and Colombia- recorded explosive growth in their number of journals (1440% to18000%) than other groups of countries.

On the other hand, the core countries-USA, UK, Germany, France, Japan and Netherlands (range 70.8% to 149%) stalled in growing their number of journals. While these countries maintained their dominance in publishing journals, it was revealed that their growth in journal publishing stalled.

Furthermore, the article addresses the mixed growth patterns in Asian countries. Japan showed relatively modest growth, India and China following a higher trajectory with about 500% increase over twenty years. South Korea experienced the highest level of growth, increasing its number of journals by about 1,000%. These diverse growth patterns reflect the influence of factors such as research investments, national policies, and regional dynamics. South Africa's (the only Sub-Saharan African country) growth was particularly remarkable, with a percentage change of 265.6%. This underscores the increasing significance of academic research and publication in these countries and regions.

Table 1b provides a comprehensive overview of the percentage changes in the number of academic journals published in the top twenty countries, as recorded in the Scopus database, during the period from 2001 to 2020. One of the most striking observations is the slow pace of increase in the number of journals from the top twenty countries. USA and UK consistently maintained their dominance with relatively modest growth rate. Iran recorded the highest percentage increase in the number of journals they produced in Scopus, with about 90% increase, followed by South Korea, Brazil and Spain. Japan (JPN) stands out as a country that faced a significant decline in journals published in Scopus within the twenty-year period, with a cumulative change of -10.9% over the entire two-decade period.

**Table 1a: Percentage Changes to the Number of Journals Published in the top Twenty Countries in Web of Science between 2001 and 2020**

| Year | USA | GBR | NLD | DEU | ESP | ITA | CHN | BRA | RUS | FRA | POL | JPN | IND | CAN | KOR | TUR | AUS | CHE | COL | ZAF | OTHERS | GLOBAL |
|---|---|---|---|---|---|---|---|---|---|---|---|---|---|---|---|---|---|---|---|---|---|---|
| 2001-2002 | 0.63 | 0.96 | -0.4 | 0 | 4.167 | 2.7 | 6.494 | 5 | -1.55 | 1.83 | 6.522 | 1.73 | 0 | -1.36 | 10.7 | 0 | -1.11 | -0.63 | 0 | -9.68 | 0.78 | 0.68 |
| 2002-2003 | 1.09 | 1.86 | 1 | 0.32 | 0 | -0.88 | 7.317 | 4.762 | 1.57 | -1.8 | 18.37 | 1.14 | 0 | -1.38 | 16.1 | 0 | 2.247 | -1.26 | 0 | 3.57 | 2.32 | 1.25 |
| 2003-2004 | 1.05 | 2.14 | -0.3 | -0.6 | 8 | 0 | 2.273 | 0 | -1.55 | -4.1 | 0 | 2.81 | 5.36 | -1.4 | 2.78 | 50 | 4.396 | 3.185 | 0 | 0 | 3.77 | 1.26 |
| 2004-2005 | 16.9 | 36.5 | 18 | 0.16 | 492.6 | 66.4 | 18.89 | 477.3 | 33.1 | 24.8 | 75.86 | 15.8 | 88.1 | 63.8 | 168 | 1333.3 | 118.9 | 13.58 | 7700 | 200 | 80.7 | 36.2 |
| 2005-2006 | 2.75 | 5.16 | 3.2 | 17.8 | 9.063 | 7.45 | 8.411 | 10.24 | 4.14 | 6.11 | 17.65 | 6.13 | 5.41 | 3.03 | 5.05 | 4.6512 | 3.365 | 2.174 | 15.385 | 9.2 | 12.3 | 4.95 |
| 2006-2007 | 3.96 | 6.01 | 4.6 | 2.72 | 18.34 | 21.3 | 30.17 | 42.86 | 20.5 | 13.3 | 40.83 | 12.4 | 29.1 | 6.3 | 29.8 | 47.778 | 25.12 | 14.89 | 22.222 | 28.4 | 32 | 11.3 |
| 2007-2008 | 7.28 | 13.2 | 9.3 | 11.2 | 15.98 | 12.7 | 17.88 | 33 | 16 | 6.67 | 20.71 | 8.3 | 19.9 | 6.32 | 15.6 | 19.549 | 11.15 | 10.65 | 22.727 | 9.84 | 19.6 | 11.7 |
| 2008-2009 | 4.55 | 7.46 | 4.2 | 7.13 | 8.559 | 13.8 | 9.551 | 13.53 | 7.72 | 5.65 | 11.27 | 3.65 | 13.3 | 5.2 | 7.69 | 7.5472 | 3.01 | 10.04 | 8.8889 | 6.72 | 10.2 | 6.66 |
| 2009-2010 | 2.14 | 4.77 | 2.1 | 3.55 | 7.115 | 10.5 | 5.128 | 9.934 | 3.4 | 2.25 | 7.048 | 3.52 | 14.1 | 3.53 | 11.3 | 7.0175 | 2.922 | 6.084 | 10.204 | 2.8 | 7.89 | 4.59 |
| 2010-2011 | 2.89 | 3.5 | 3.5 | 4.72 | 6.284 | 5.19 | 6.341 | 6.928 | 6.93 | 2.48 | 4.938 | 1.02 | 7.69 | 4.1 | 6.42 | 8.7432 | 2.524 | 15.41 | 6.1728 | 3.4 | 6.21 | 4.25 |
| 2011-2012 | 2.11 | 3.41 | 3.8 | 3.89 | 7.601 | 3.29 | 12.39 | 6.479 | 3.41 | 1.88 | 8.235 | 2.36 | 6.35 | 5.57 | 8.54 | 6.0302 | 0.308 | 7.143 | 4.6512 | 1.32 | 5.89 | 3.91 |
| 2012-2013 | 2.36 | 3.12 | 4.4 | 4.63 | 5.024 | 2.92 | 11.84 | 3.175 | 3.63 | -0.3 | 7.246 | 0.33 | 5.22 | 1.55 | 7.41 | 0.9479 | 2.454 | 11.3 | 3.8889 | 1.3 | 3.99 | 3.27 |
| 2013-2014 | 1.04 | 2.15 | 3 | 1.13 | 3.438 | 1.8 | 5.109 | 1.538 | 2.23 | -1.6 | 6.081 | 1.64 | 0.35 | 0.61 | 5.17 | 5.6338 | 2.994 | 5.469 | -1.6043 | -0.64 | 2.33 | 1.98 |
| 2014-2015 | 1.21 | 2.12 | 2.5 | 1.68 | 2.746 | 8.61 | 4.861 | 1.768 | 3.74 | 2.15 | 3.503 | 3.55 | 6.71 | 2.13 | 2.46 | 1.3333 | 3.488 | 6.667 | 0.5435 | 2.58 | 2.28 | 2.33 |
| 2015-2016 | 1.35 | 1.55 | 1.1 | 2.11 | 3.376 | 1.63 | 6.291 | -0.248 | 5.11 | 0.26 | 0.308 | 1.87 | 1.99 | -0.89 | 6.4 | 4.386 | -2.53 | 3.704 | 0 | 3.77 | 1.64 | 1.65 |
| 2016-2017 | 1.46 | 1.15 | 1.5 | 1.79 | 0.816 | 0.69 | 8.723 | 2.736 | 7.14 | -0.3 | 4.601 | 4.89 | 0.97 | 0.3 | 3.76 | 5.8824 | -1.15 | 5.804 | 1.0811 | -0.61 | 3.26 | 1.94 |
| 2017-2018 | 0.69 | 0.85 | 0.9 | 0.88 | 0 | -0.46 | 7.163 | 1.211 | 2.4 | -1.3 | 2.053 | 2.33 | 5.14 | 1.5 | 0.36 | 0.7937 | -1.17 | 4.852 | -2.139 | -1.22 | 2.01 | 1.14 |
| 2018-2019 | 1.61 | 1.96 | 2.9 | 1.14 | -0.94 | 0.69 | 10.43 | 0 | 1.3 | 0.8 | 1.149 | 1.14 | 7.34 | -0.88 | 3.61 | -3.543 | -1.18 | 8.451 | -0.5464 | 1.23 | 0.62 | 1.66 |
| 2019-2020 | 0.39 | 0.69 | 1.2 | 0.35 | 0.817 | 0.23 | 4.843 | 0.478 | 0 | -0.8 | 1.136 | -0.28 | 0.57 | 0.6 | 3.48 | 5.3061 | -1.79 | 1.484 | -0.5495 | -0.61 | 1.3 | 0.77 |
| **2001-2020** | **70.8** | **149** | **90** | **85.6** | **1442** | **297** | **462.3** | **2000** | **202** | **71.2** | **673.9** | **105** | **530** | **130** | **961** | **6350** | **265.6** | **241.9** | **18000** | **426** | **444** | **156** |

**Table 1b: Percentage Changes to the Number of Journals Published in the top Twenty Countries in Scopus between 2001 and 2020**

| Year | USA | GBR | NLD | DEU | CHN | CHE | ESP | ITA | FRA | IND | POL | JPN | RUS | BRA | KOR | CAN | IRN | TUR | CZE | AUS | OTHERS | GLOBAL |
|---|---|---|---|---|---|---|---|---|---|---|---|---|---|---|---|---|---|---|---|---|---|---|
| 2001-2002 | 5.3 | 3.3 | 0.0 | 5.7 | -0.5 | 1.0 | 10.5 | 17.1 | 13.4 | 5.0 | 6.7 | 0.2 | 1.1 | 9.2 | 0.0 | 9.0 | 28.6 | 18.0 | 6.9 | 7.5 | 6.8 | 4.7 |
| 2002-2003 | -0.1 | 2.2 | 2.0 | 2.2 | 4.5 | 0.0 | -0.9 | -2.0 | 0.0 | 1.9 | 1.7 | 0.0 | 0.5 | 2.8 | 7.8 | -2.9 | 22.2 | 11.1 | 1.1 | -4.2 | 1.2 | 1.1 |
| 2003-2004 | 0.7 | 3.3 | 2.6 | 1.9 | 4.5 | -1.0 | 2.7 | 0.6 | 4.1 | 2.3 | 1.7 | -0.8 | -1.1 | 2.7 | 3.6 | 0.0 | 63.6 | 0.0 | 1.1 | -0.7 | 1.4 | 1.9 |
| 2004-2005 | 1.0 | 2.5 | 2.3 | 3.9 | 8.0 | -0.3 | 9.3 | 0.3 | 1.7 | 5.5 | 7.7 | 0.0 | 2.7 | 3.5 | 7.0 | -0.8 | 38.9 | 10.0 | 5.3 | 2.2 | 4.4 | 2.4 |
| 2005-2006 | 0.9 | 3.0 | 1.5 | 2.8 | 6.2 | 2.4 | 5.3 | 1.2 | 2.4 | 7.4 | 9.2 | 1.0 | -0.5 | 41.9 | 9.8 | -0.4 | 32.0 | 12.5 | -2.0 | 5.0 | 7.5 | 3.0 |
| 2006-2007 | 1.0 | 1.9 | 3.4 | 1.2 | -4.2 | 8.1 | 6.2 | 4.4 | 1.4 | 6.9 | 10.8 | -1.2 | -0.5 | 21.1 | 32.8 | 0.9 | 21.2 | 2.0 | 11.2 | 3.4 | 11.2 | 2.9 |
| 2007-2008 | 1.9 | 3.3 | 2.9 | 3.3 | 3.4 | 11.5 | 14.9 | 4.5 | 3.7 | 9.8 | 15.7 | 2.9 | -16.3 | 14.9 | 34.8 | 0.9 | 42.5 | 21.8 | 5.5 | 17.2 | 18.8 | 5.1 |
| 2008-2009 | 3.9 | 2.9 | 2.3 | 5.1 | 3.1 | 9.5 | 10.1 | 4.8 | 1.1 | 21.7 | 10.6 | 1.0 | -33.3 | 10.0 | 11.7 | 7.6 | 21.1 | 26.0 | 7.0 | 13.6 | 13.8 | 5.1 |
| 2009-2010 | 2.8 | 3.5 | 2.1 | 2.9 | 1.5 | 9.9 | 6.9 | 5.4 | 2.4 | 15.3 | 1.7 | 0.4 | -1.9 | 9.1 | 17.9 | 0.4 | 29.0 | 9.7 | 4.1 | 6.0 | 7.5 | 4.1 |
| 2010-2011 | 2.2 | 1.9 | 6.0 | 3.3 | 1.6 | 9.3 | 16.9 | 6.8 | 5.7 | 14.5 | 6.5 | 0.0 | 11.5 | 8.3 | 14.6 | 4.7 | 22.5 | 9.4 | 14.1 | 7.5 | 11.5 | 4.8 |
| 2011-2012 | 0.9 | 1.8 | 5.3 | 2.2 | 1.4 | 6.6 | 11.0 | 4.1 | 1.9 | 7.1 | 5.8 | 0.6 | 11.2 | 8.0 | 6.1 | 3.0 | 11.9 | 2.7 | 4.1 | 0.4 | 5.1 | 2.9 |
| 2012-2013 | 1.6 | 2.0 | 1.2 | 2.4 | 5.7 | 8.8 | 3.5 | 1.3 | 1.7 | 0.4 | 4.3 | -0.4 | 15.5 | 2.5 | 4.2 | 1.1 | 8.2 | 3.7 | 5.3 | -0.4 | 3.6 | 2.4 |
| 2013-2014 | 1.4 | 2.6 | 4.2 | 2.4 | 1.3 | 8.4 | 2.6 | 3.0 | -0.7 | 3.4 | 2.5 | -0.6 | 10.7 | 1.8 | 8.0 | 0.4 | 7.6 | 2.5 | 7.5 | -0.4 | 3.0 | 2.5 |
| 2014-2015 | 0.8 | 2.0 | 3.0 | 1.7 | 1.3 | 10.0 | 2.9 | 3.1 | -0.8 | 0.4 | 2.7 | 0.4 | 15.8 | 2.4 | 6.0 | 0.7 | 8.5 | 2.0 | 2.3 | 0.0 | 2.7 | 2.1 |
| 2015-2016 | 0.9 | 1.0 | 1.6 | 0.1 | 6.9 | 4.1 | 4.2 | 2.8 | 0.2 | 2.7 | 8.4 | -0.8 | 34.0 | 1.7 | 3.5 | 0.0 | 7.8 | 6.8 | 3.4 | -0.4 | 3.1 | 2.0 |
| 2016-2017 | 1.0 | 0.9 | -0.8 | 1.1 | 1.1 | 3.7 | 5.5 | 3.6 | -1.7 | -0.7 | 2.4 | -2.0 | 24.6 | 6.0 | 8.0 | 1.8 | 8.4 | 3.6 | 6.6 | -1.3 | 3.9 | 1.8 |
| 2017-2018 | 1.9 | 2.5 | 0.4 | 2.8 | 4.7 | 1.8 | 9.5 | 5.7 | 3.3 | 5.7 | 7.6 | 4.5 | 17.9 | 5.6 | 10.9 | 8.4 | 10.6 | 11.4 | 5.7 | 6.3 | 5.4 | 3.6 |
| 2018-2019 | 0.2 | 0.6 | 0.6 | 1.1 | 0.6 | 0.1 | 4.4 | 3.8 | -0.7 | 2.1 | 6.2 | -2.7 | 13.8 | 2.5 | 4.6 | -1.0 | 14.6 | 0.8 | 3.9 | 2.1 | 3.4 | 1.4 |
| 2019-2020 | -4.5 | -1.2 | -1.2 | -5.5 | 4.7 | -1.7 | 4.8 | -1.4 | -6.5 | -13.6 | 2.1 | -11.8 | 0.9 | 1.0 | 5.1 | -4.5 | 7.5 | -5.8 | 0.0 | -12.8 | -2.2 | -2.6 |
| **2001-2020** | **21.0** | **32.7** | **32.0** | **32.8** | **41.9** | **58.7** | **71.2** | **48.6** | **26.7** | **59.7** | **66.7** | **-10.9** | **56.9** | **76.0** | **83.7** | **24.1** | **97.1** | **74.8** | **59.2** | **37.3** | **65.7** | **39.5** |

*Citations received by Journals from the top twenty countries in Web of Science and Scopus between 2001 to 2020*

Figures 3a and 3b present data on the journal visibility of the top twenty countries in Web of Science and Scopus, respectively, spanning the years 2001 to 2020. The trend in journal citation share varies among the countries. The United States, traditionally the predominant contributor to journals in Web of Science, experienced a substantial decline in its journal citation share over the period from 2001 to 2020. Starting with the highest journal citation share of 58.64% in 2001, the U.S. contribution dwindled to 35.39% by 2020, indicating a significant decrease.

Contrary to the USA's decline, most other countries within and outside the top twenty in Web of Science increased the share of citations their journals received. For instance, the United Kingdom's journal citation share increased from 21.98% in 2001 to 28.54% in 2020, indicating that UK journals were cited more frequently in 2020 than in 2001. Similarly, Spain, Italy, Brazil, China, Poland, India, South Korea, Turkey, South Africa, and Australia demonstrated an increase in their share of citations received by all the journals in Web of Science. Switzerland had a relatively remarkable increase in the share of citations its journals in Web of Science received, from 1.40% in 2001 to 9.93% in 2020, despite having only a 1.88% to 2.50% share of all the journals in Web of Science. Notably, China saw a significant rise in the share of citations its journals in Web of Science received, from 0.21% in 2001 to 2.18% in 2020, even though China had only a 0.90 to 1.98% share of all the journals.

The combined share of the USA and the UK journal citations in the Web of Science declined from 80.66% in 2001 to 63.93% in 2020. Furthermore, the journal citation shares of the top four countries—USA, UK, Netherlands, and Germany—decreased from 93.05% in 2001 to 79.89% in 2020. Despite the decrease in the journal citation share of the UK, USA, Germany, and the Netherlands, these countries are overrepresented in citations received compared to journals from other countries. For instance, these four countries had 59.92% of all the journals in Web of Science in 2020 but received 79.89% of the citations received by all the journals in Web of Science. Concurrently, the journal citation shares of countries outside the top twenty increased from 1.98 to 3.20 in 2020, reflecting a changing landscape in the distribution of the visibility of globally indexed journals in Web of Science. However, countries from outside the top twenty accounted for 12.79% of all the journals in Web of Science but received only 3.2% of all the citations.

Figure 2b presents the data on the share of the top twenty countries in Scopus journals globally between 2001 and 2020. Similar to Web of Science, the share of journal citations received by the journals from the USA in Scopus consistently reduced from 2001 to 2020. In contrast, the share of citations received by journals from other countries apart from the USA, except for Germany, Russia, Japan, and Canada, consistently increased from 2001 to 2020. The United States' contribution to the global journal citation share declined from 2001 to 2020, starting with the largest share of 55.19% in 2001, and saw the highest decline, with its contribution dropping to 34.54% in 2020. The journal citation shares of the other big four countries in Scopus either consistently increased or largely remained unchanged. For instance, the UK increased its journal citation share from 26.20% in 2001 to 32.67%, in 2020, almost at par with the USA. The Netherlands' journal citation shares in Scopus increased, while Germany's largely remained the same. Turkey, Iran, Brazil, and India saw the highest increase in their Scopus journal citation share between 2001 and 2020. Iran increased its journal citation share in Scopus from less than 0.01% to 0.16%. Turkey also increased its Scopus journal citation share from 0.01% to 0.14%. Brazil also increased its share of the journal citation share in Scopus from 0.05% to 0.32 within the twenty-year period.

The share of the visibility of Scopus journals from the USA and UK combined reduced from 81.39% in 2001 to 67.21% in 2020. The declining trend continues for the combination of the big four countries—USA, UK, Netherlands, and Germany—with the percentage of journal citation reduced from 94.65% in 2001 to 85.18% in 2020. Whereas the journal citation share of the countries outside the top twenty increased from 1.59% in 2001 to 3.40% in 2020.

**Figure 3a: Citations received by Web of Science Journals Published in the top twenty countries between 2001 to 2020**

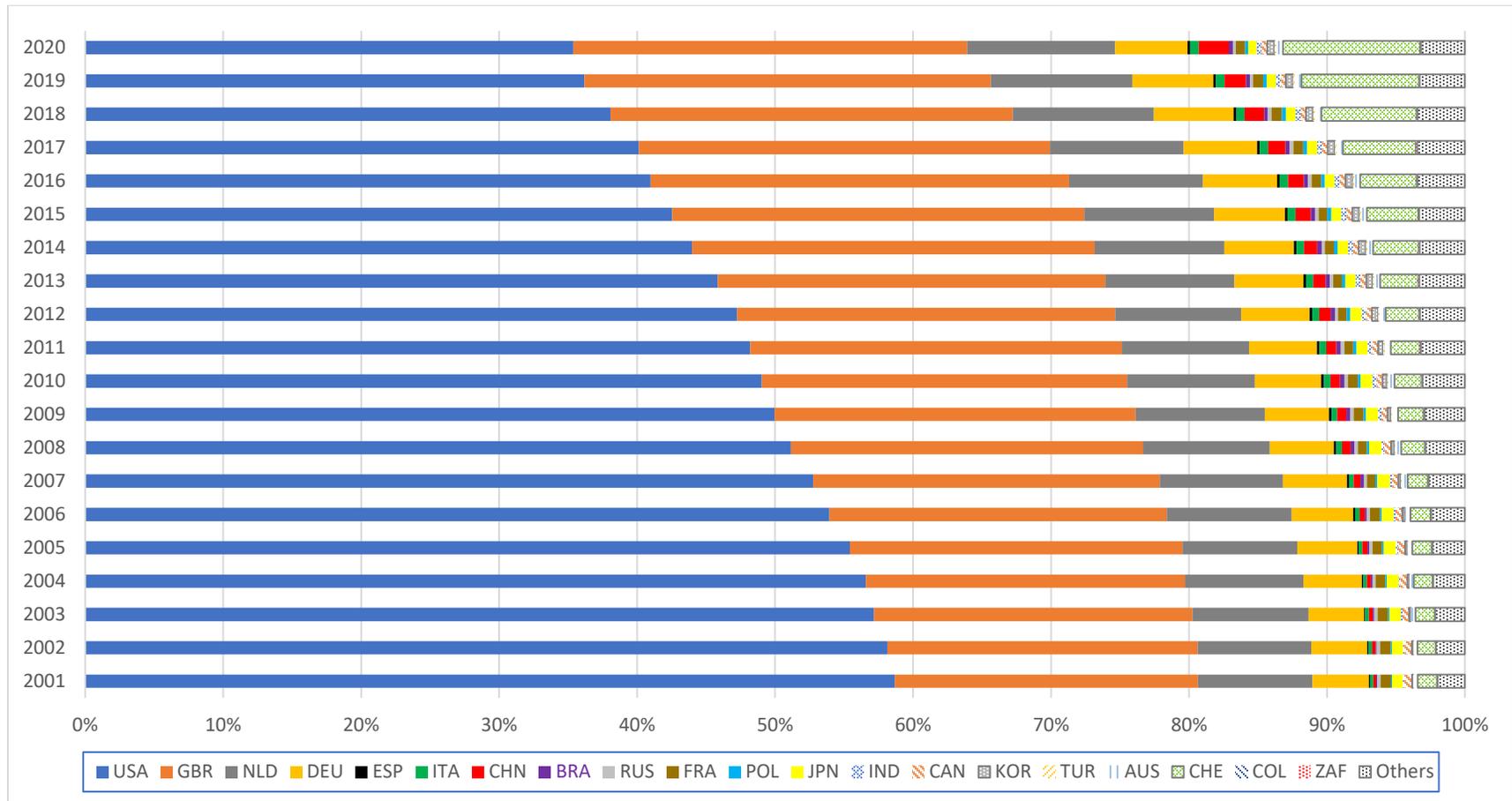

**Figure 3b: Citations received by Scopus Journals Published in the top twenty countries between 2001 to 2020**

| Year | USA | GBR | NLD | DEU | CHN | CHE | ESP | ITA | FRA | IND | POL | JPN | RUS | BRA | KOR | CAN | IRN | TUR | CZE | AUS | OTHERS |
|---|---|---|---|---|---|---|---|---|---|---|---|---|---|---|---|---|---|---|---|---|---|

Stacked bar chart showing the percentage share of countries (USA, GBR, NLD, DEU, CHN, CHE, ESP, ITA, FRA, IND, POL, JPN, RUS, BRA, KOR, CAN, IRN, TUR, CZE, AUS, OTHERS) from 2001 to 2020.

## Web of Science and Scopus Journal Prestige of the top twenty Countries between 2001 and 2020

Figures 5a and 5b presents the Web of Science and Scopus journal prestige, respectively. Figure 5a shows the average impact factor of journals from the top twenty countries in Web of Science between 2001 and 2020. The trends in Web of Science suggests that journal prestige is flat or declined in the twenty years, except for China, that increased its journal prestige.

United Kingdom consistently produced journals with the highest average Impact Factor in Web of Science from 2001 to 2010 and maintained second position afterwards. By 2010, China overtook UK as the country with the highest average Impact Factor in Web of Science and maintained the leading spot until 2020. The USA consistently has the third highest average Impact Factor among all the countries, ranging from 4.15 to 4.48, closely followed by Netherlands. China has the most remarkable growth in average impact factor, from 3.43 in 2003 to 6.04 in 2020. In contrast, The Netherlands has the highest decline in impact factor, from 4.33 in 2002 to 4.01 in 2020.

Average impact factor of fifty percent of countries (Spain, Brazil, France, Poland, Japan, India, Canada, Turkey, Australia and South Africa) ranged between 1.0 and 3.0 in 2020. Colombia and Most of the other countries have stable or slight changes in impact factor, with some fluctuations over the years. The only two countries with average impact factor lower than one in 2020 are Columbia and Russia; their average impact factor ranged from 0.79 to 1.92 and 0.86 to 0.98, respectively.

Figure 5b shows the Scopus Scimago Journal Rank of the top twenty countries between 2001 and 2020. USA has the highest rank in every year, and its rank has increased from 0.71 in 2001 to 0.95 in 2020. USA is followed by UK and Netherlands which also have high and stable ranks throughout the years. China has shown the most remarkable growth in its rank, from 0.10 in 2001 to 0.31 in 2020, surpassing many other countries. It is the only country that has more than doubled its rank in this period. Iran has the second fastest growing average Scimago journal rank, from 0.08 in 2001 to 0.23 in 2020, followed by South Korea and Brazil. Switzerland is also remarkable growing SJR from 0.39 in 2001 to 0.64 in 2020, and surpassing German at number four position.

Russia has the lowest rank among the top twenty countries, and its rank has fluctuated between 0.10 and 0.19. It has not shown any consistent improvement over the years. Average SJR of 60% of the top twenty countries ranged between two and 3.5.

**Figure 4a: Web of Science Journal Prestige of the top twenty countries between 2001 and 2020**

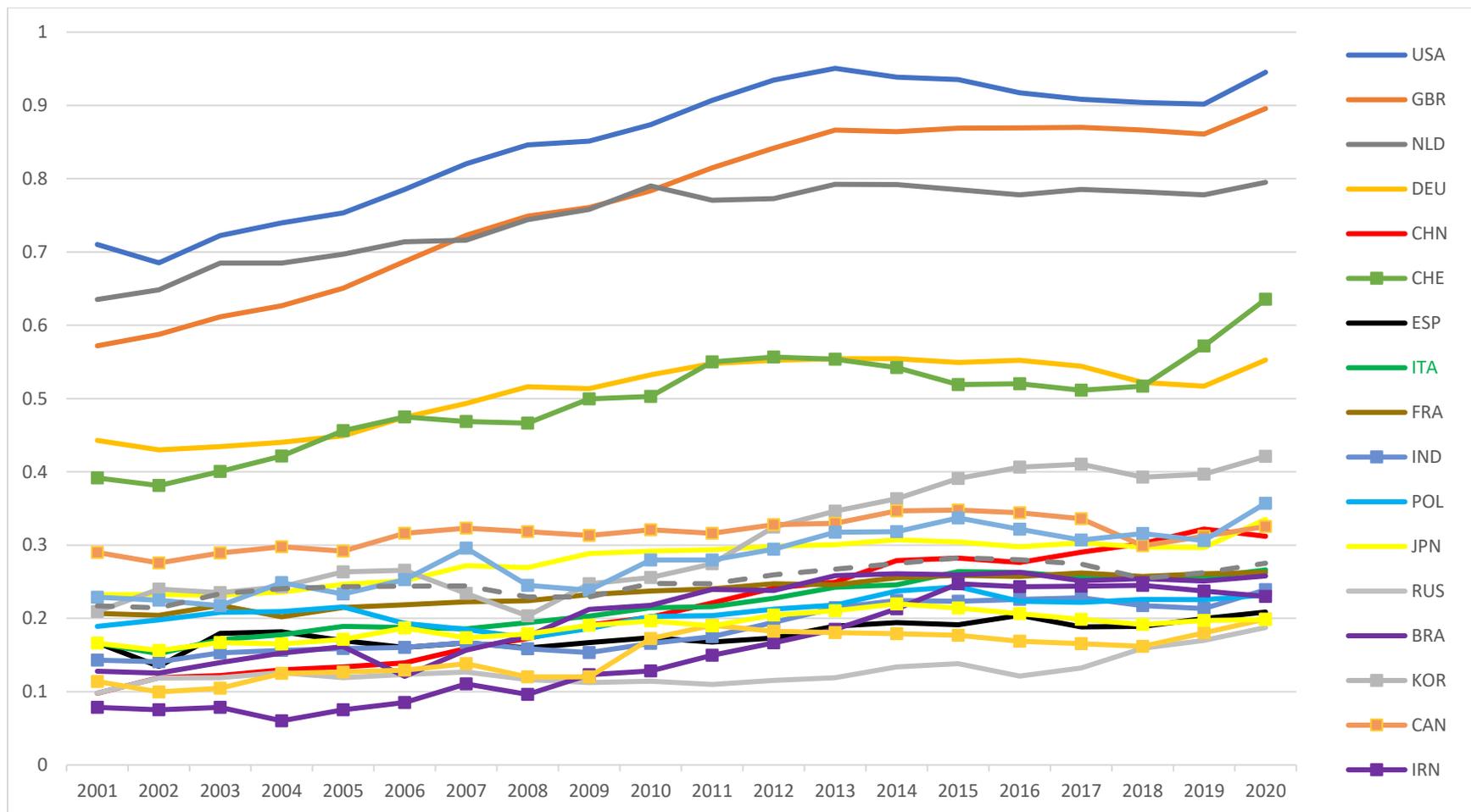

**Figure 4b: Scopus Journal Prestige of the top twenty countries between 2001 and 2020**

## 4. Discussion and Recommendation

In summary, Web of Science and Scopus journal coverage, visibility and prestige among the top twenty countries and others outside the top twenty delineate diverse academic publishing trends in the two foremost academic databases globally. The USA and Great Britain maintain their global journal publishing and visibility leadership in Web of Science and Scopus, though their influence is waning. The emergence of China in 2005 as the leader in the most prestigious journals in Web of Science is a break from the norm of having USA and the UK at the forefront, though the two countries produced the most prestigious journals in Scopus. Europe and Asia exhibit mixed growth patterns, illustrating the complexities of research ecosystems and national policies. The emergence of South Africa and Colombia as key players in academic publishing signals a changing landscape in the global scholarly communication arena. In the Scopus database, the USA and Great Britain continue their global leadership with impressive growth, while Iran, the Czech Republic, and Japan experience comparatively modest or negative changes in academic publishing. These observations underscore the dynamic and complex nature of scholarly communication influenced by various factors, including research investments, national policies, and regional dynamics, prompting the need for further research to understand underlying causes and consequences.

The observed trends in the Web of Science and Scopus databases provide a nuanced understanding of the evolving landscape of global scholarly contributions. In the Web of Science, European dominance is evident, with the United Kingdom, Germany, and the Netherlands leading, reflecting a historical trend. Southeastern Asia and the Americas play significant roles, while Sub-Saharan Africa and Oceania exhibit more restrained contributions. Substantial growth variations are observed in the "OTHERS" category, highlighting dynamism outside the top twenty. Spain's ascent into the top five and the sustained USA dominance characterize the Web of Science landscape, emphasizing the complexity of global scholarly contributions.

Contrastingly, Scopus portrays a different trajectory with European nations at the forefront, diverging from Web of Science trends. Japan's journal reduction and Iran's notable growth outside the top twenty underscore the distinctiveness of Scopus's evolution, emphasizing the nuanced dynamics of these databases, each presenting a unique snapshot of global scholarly contributions.

Examining specific country contributions within the Web of Science, the historically primary contributor, the United States, witnessed a significant decline in its share from 40.77% in 2001 to 27.16% in 2020. Germany, France, Japan, and the Netherlands also experienced decreases, while the United Kingdom maintained stability. Conversely, Spain, Italy, Brazil, China, Poland, India, Korea, Turkey, South Africa, and Australia exhibited noteworthy increases in global journal shares. Spanish-speaking countries, particularly Spain and Colombia, demonstrated substantial rises, exemplified by Colombia's surge from 0.01% in 2001 to 0.83% in 2020. Concurrently, countries outside Europe and non-English-speaking regions increased shares, contributing to the decline in the combined Web of Science journal share of the USA and UK from 62.62% in 2001 to 48.43% in 2020, with an increase in the share of countries outside the top twenty from 6.02% to 12.78%.

Examining the citation landscape within the Web of Science, the United States exhibited a notable decline in its journal citation share from 58.64% in 2001 to 35.39% in 2020. Conversely, most other countries, within and outside the top twenty, experienced increased citation shares. The United Kingdom's journal citation share rose from 21.98% to 28.54%, signaling increased citation frequency in 2020. Spain, Italy, Brazil, China, Poland, India, South Korea, Turkey, South Africa, and Australia also demonstrated heightened citation shares. Despite the decline in the combined share of USA and UK journal citations from 80.66% in 2001 to 63.93% in 2020, these countries, along with Germany and the Netherlands, retained an overrepresented position, receiving 79.89% of all citations despite constituting 59.92% of journals. Simultaneously, the journal citation shares of countries outside the top twenty increased from 1.98% to 3.20% in 2020, reflecting shifting dynamics in the global distribution of citations.

A parallel pattern was observed in Scopus, with the USA's citation share decreasing while other countries, particularly Turkey, Iran, Brazil, and India, saw significant increases in their Scopus journal citation shares between 2001 and 2020. The share of Scopus journals from the USA and UK combined reduced from 81.39% in 2001 to 67.21% in 2020, aligning with the declining trend in the big four countries (USA, UK, Netherlands, and Germany) from 94.65% to 85.18%. Meanwhile, the citation share of countries outside the top twenty in Scopus increased from 1.59% in 2001 to 3.40% in 2020, indicating a noteworthy diversification in the global citation landscape.

These findings contribute to the existing literature by highlighting the intricate dynamics of global scholarly contributions as captured by Web of Science and Scopus. The observed shifts in contributions and citations underscore the need for a nuanced understanding of the scholarly landscape, considering the specific characteristics and trajectories of each database. This discussion prompts further exploration into the factors influencing these trends and their implications for the assessment of scholarly impact on a global scale. The decline of the USA is notable and could reflect changes in the global research landscape, with other countries catching up. This finding suggests potential obstacles to research and scholarly communication in the Japanese context, which warrant further investigation.

*Limitations of the Study and Suggestions for Further Studies*

One limitation arises from the data collection process, as the study relies on Web of Science and Scopus journal lists obtained from their respective platforms. The exclusion of certain source types, such as conference proceedings, book series, and books, may lead to an incomplete representation of scholarly output. Additionally, the decision to exclude Crimea from the analysis due to its non-recognition as a country or territory by the United Nations introduces a potential geographical bias. The method employed to resolve missing journal country information by consulting external sources like Scimagor, ISSN portal, DOAJ, Worldcat, and journal websites could introduce variability in the accuracy of the data, as determinations were based on external references. The use of Scopus data from Scimagojr.com, a third party source, could constitute a limitation. Furthermore, the collation of records for England, Scotland, Wales, and Northern Ireland under the United Kingdom category might oversimplify the distinct scholarly contributions of these regions. Lastly, the ranking of South Africa over Iran in the Web of Science analysis is based on the number of journals, and this decision could impact the representation of these countries in the final results. Despite these limitations, the study provides valuable insights into the trends and dynamics of global scholarly contributions within the specified constraints.

**Acknowledgement**

ChatGPT was used for manuscript development.

**Author contribution**

Conceptualization: [Toluwase Asubiaro]; Data Curation: [Toluwase Asubiaro], [Toluwase Asubiaro]; Methodology: [Toluwase Asubiaro]; Formal analysis and investigation: [Toluwase Asubiaro], [Toluwase Asubiaro]; Writing - original draft preparation: [Toluwase Asubiaro, ], []; Writing - review and editing: [Toluwase Asubiaro], [], [Toluwase Asubiaro]

**Competing interest statements**

The authors have no relevant financial or non-financial interests to disclose.

**Funding information**

None

**Data availability statements**

Data available on request.